# Flexible Authentication Technique for Ubiquitous Wireless Communication using Passport and Visa Tokens

Abdullah M.Almuhaideb, Mohammed A. Alhabeeb, Phu D.Le, and Bala Srinivasan

**Abstract**— The development of mobile devices (CPU, memory, and storage) and the introduction of mobile networks (Ad-Hoc, Wi-Fi, WiMAX, and 3.5G) have opened new opportunities for next generation of mobile services. It becomes more convenience and desirable for mobile internet users to be connected everywhere. However, ubiquitous mobile access connectivity faces interoperation issues between wireless network providers and wireless network technologies. Although mobile users would like to get as many services as possible while they travel, there is a lack of technology to identify visited users in current foreign network authentication systems. This challenge lies in the fact that a foreign network provider does not initially have the authentication credentials of a mobile user. Existing approaches use roaming agreement to exchange authentication information between home network and foreign network. This paper proposes a roaming agreement-less approach designed based on our ubiquitous mobile access model. Our approach consist of two tokens, Passport (identification token) and Visa (authorisation token) to provide the mobile user with a flexible authentication method to access foreign network services. The security analysis indicates that our proposal is more suitable for ubiquitous mobile communication especially in roaming agreement-less environment.

——————————— ◆ ———————————

## 1 INTRODUCTION

THE development of mobile devices has grown significantly over the last decade from a simple mobile phone to a pocket size-computing device with the capability to access the Internet via various wireless systems such as Wi-Fi and 3.5G networks. The advanced capabilities of mobile devices allow mobile users (MUs) to pay for products, surf the internet, buy and sell stocks, transfer money and manage bank accounts on the move without being restricted to a specific location. This fact keeps attracting many MUs to be connected wirelessly. It is estimated that half the world's population now pays to use mobile devices[1].

A MU always asks for a higher speed at lower prices, and demands to be "Always Best Connected" [2]. The MU also wants a ubiquitous wireless coverage to network resources from anywhere, anytime. Yet, it is hard to achieve both high data rate and wide coverage at once. For a smaller coverage, it is easier to provide higher data rates. For instance, a 3.5G networks have a wider coverage but slower speeds; while Wi-Fi networks have higher speeds but smaller coverage. A key challenge in such heterogeneous networks is the possibility of roaming to administrative domains with which a MU's home domain does not have a pre-established roaming agreement [3]. Therefore ubiquitous wireless network coverage with high data rates is not feasible with a single technology and a single wireless provider. A heterogeneous wireless network will be composed of wireless networks of multiple technologies operated by multiple network providers.

Most of the current mobile devices are built with multiple wireless interfaces. They have built-in chipsets for IEEE 802.11 based Wireless LAN (WLAN) and interfaces for data connectivity using cellular networks. Nowadays, university campuses and company offices are supported by WLAN allowing their students or employees to have free access to the wireless networks. Hotspot operators offer wireless Internet in public places like cafés, restaurants, hotels and airports. A Wi-Fi community called FON has more than 250,000 hotspots worldwide [4], operated by individuals sharing their home Wi-Fi connection with other FON community members. An increasing number of wireless technologies and growing number of wireless providers of different sizes have in fact built a heterogeneous wireless network towards a worldwide coverage.

**Problem Statement.** This growing number of wireless technologies and providers, as well as users' increasing need and desire to be connected and reached at all times, call for the development of ubiquitous wireless access. However, authenticating unknown users by foreign network (FN) providers is still a challenge.

Fig.1 shows that a MU cannot get network service from FN if there is no roaming agreement with the MU's

————————————————

- *A.M Almuhaideb is with the Faculty of Information Technology, Monash University, Melbourne, Australia.*
- *M.A. Alhabeeb is with the Faculty of Information Technology, Monash University, Melbourne, Australia.*
- *P.D Le is with the Faculty of Information Technology, Monash University , Melbourne, Australia.*
- *B.Srinivasan is with the Faculty of Information Technology, Monash University, Melbourne, Australia.*

home network (HN) for verification process. Traditionally, for MUs to be able to roam into FNs, the FN and the MU's HN must trust each other and have a roaming agreement established beforehand. However, HN cannot establish service agreement with all FNs. Therefore, MUs will not be able to get the service unless they could identify themselves to the FN.

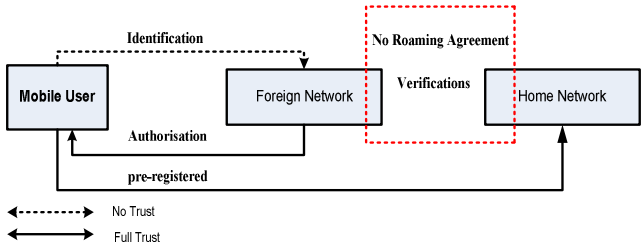

Fig. 1. Roaming agreement-less challenge.

**Our Approach and Contributions.** This research proposes the Passport-Visa technique based on our ubiquitous mobile access model. The "Passport" is an authentication token that issued by identity provider (IdP: certificate authority or HN) to MU in order to identify and verify MU identity. The Passport in itself does not grant any access, but provides a unique binding between an identifier and the subject. The "Visa" is an authorisation token that granted to a MU via a FN. The Visa token can be used as an access control to ban individual users. The contribution of this proposed approach is that:

- It is wireless technology independence: it is not feasible to achieve ubiquitous mobile access with single wireless technology. Therefore, the authentication solution should enable access to the core network regardless of the wireless technology. The proposed authentication solution is not designed for specific underlying wireless technology. It is aimed to be designed at the network layer of the OSI to avoid the differences in the link and physical layer.
- Support direct negotiation: MUs should be able to choose and select the appropriate network based on direct negotiation of services and authentication. The proposed solution supports direct negotiate with the MU not with the IdP, which will increase the satisfaction of the user.
- Support Open Marketplace (Home Network Independent): MU should be able to access FN without the control of the HN. MUs can get the benefit of HN partners and more. They could get more network services in area not covered by HN's partners with full freedom of choice.
- It is roaming agreement independence: A current solution by a visited MU's cellular HN is based on a roaming agreement to establish trust with other cellular network providers to extend services for their users. It is not likely to set up formal roaming agreements with every possible provider by MU's HN. Our approach does not depend on roaming agreement between FN provider and IdP. Alternatively, FN provider use negotiation and trust decision to either authorize the MU or not.
- The MU holds a global identification token (Passport) to be authenticated everywhere by FNs.
- The FN service owner grants the network service authorisation token (Visa) to MU, which support the access control role of FN.
- FNs can use our technique to authenticate visited MU, although service level agreement between FN and MU's HN does not exist. As our approach does not rely on static agreement in providing services, Passport-Visa technique will offer a flexible way to authenticate and authorise MU.
- The proposed technique is lightweight for mobile devices as their side employs only symmetric keys, while the more powerful servers side used asymmetric keys.

**Organization of this paper.** The rest of this paper is structured as follows. This paper will start with an overview of the ubiquitous mobile access model, flexible authentication architecture and the Passport and Visa protocols (Section 2), where Passport acquisition, Visa acquisition, mobile service provision, Passport and Visa revocation are illustrated. We will then demonstrate the security analysis (Section 3), which followed by a review of existing approaches to the problem (Section 4).Finally, our conclusion of this paper will be presented (Section 5).

## 2 THE PROPOSED MODEL

### 2.1 Ubiquitous Mobile Access Model

To achieve ubiquitous wireless access, MUs should be able to have a direct negotiation with potential FNs regarding service provision. IdPs are required to verify the MU's identity and credentials. There should be more flexible ways to establish trust without relying on service agreements. Figure 2 below illustrates the proposed model based on direct negotiation and flexible trust. MU is pre-registered with IdP to get identification token. In this model, MUs are able to negotiate directly with potential FN providers to get the authorization token. Also, FN providers are able to communicate directly with potential MUs and make trust decision whether or not to provide network service. For a FN provider to trust MU, IdP is used to verify the claimed identity of the MU. Certificate authority can be employed to establish trust with this IdP. The relationships between engaging parties can be described as follow:

- Trust: there are three type of trust in the proposed model. The first type is "No Trust", this type can exist between MU and potential foreign network provider as a first step of communication. The second type is "Partial Trust", this type exists between FN and IdP. The term partial trust means that there is no roaming agreement between these two entities. The third type is the full trust and this one exists between MU and IdP (after the registration process) and FN (after the authentication process).

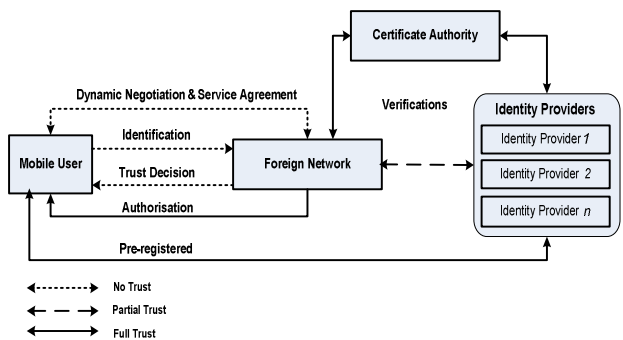

Fig. 2. The Proposed Model.

- Negotiation: a MU negotiates with prospective FN providers' the available services.
- Identification & Verification: Identification is the process of receiving credential from MU, and verification is the process of checking credential with the user's IdP.
- Authorization: after verifying potential customer identity, the FN provider decides whether to provide the service or not, based on its policy on trust decision. The MU may need to reduce the level of service to the appropriate authentication credentials level s/he may be able to provide. After the first successful authentication, MU could access the FN provider resources using the issued authorization token without any further communication with IdP.

## 2.2 Flexible Authentication Architecture

In order to achieve a flexible way to authenticate and trust a MU, a Flexible Authentication Architecture (FAA) is introduced, it is based on three components namely negotiation, trust, and policy managers, as shown in Figure 19. The following subsections introduce these three components.

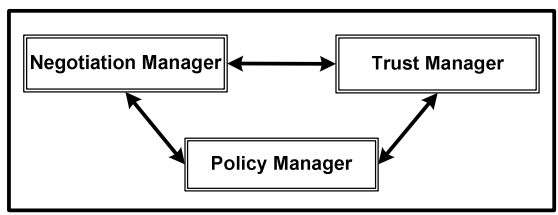

Fig. 3. Flexible Authentication Architecture.

### 2.2.1 Trust Manager

Trust is an essential component required for cooperation between ubiquitous mobile access entities. Without prior agreements, establishing trust among parties is the driving factor for inter-working. Without trust, there is no assurance that services will be delivered and paid for. Trust Manager will be responsible for making the trust decision in collaboration with Negotiation Manager based on the rules that are controlled by the Policy manager. Existing trust models will be used by Trust Manager in making trust decision. An overview of the trust management, and decision concept is described below.

**Trust Management:** Trust is a relationship between two

parties, and trust management can be described from two points of view [5]: Firstly, it enables relying parties to make assessments and decisions regarding the dependability of potential remote parties, which may involve transaction risk. Secondly, it allows parties to increase the level of trust and the reliability of their own, and appropriately represent and communicate them to other players.

**Trust decision:** Trust decision can be defined as the level to which a given party is ready to depend on another party in a given situation with a feeling of security to some extent, although negative consequences are possible [5]. FN providers play as a relying party that depends on IdPs to trust MU. However, FN providers require an approach to trust IdPs in the same way. As the state of trust can change, the relying party uses assessment parameters to make the trust decision. Trust decision can be affected by a number of factors such as reputation, risk, importance, and tolerance level [6]. Other factors can also be considered such as utility (of possible outcomes), environmental factors (law enforcement, contracts, security mechanisms etc.) and risk attitude (risk taking, risk averse, etc.) [5].

### 2.2.2 Negotiation Manager

Negotiation is an essential part in doing business as one negotiates in buying and selling. Negotiation is the protocol used by engaging parties to reach an agreement that meet every one's interests, and can be done using a simple request/response protocol. Negotiation is needed when the selling service can vary along several parameters, and when the provider is willing to offer a competitive price. The parties want to agree on a number of values at which an exchange can take place. Negotiation Manager enables engaging parties to negotiate to achieve the following [7] goals:

- Establish Trust: The two parties negotiate and agree on methods for identification and authentication to establish trust. With the mutual trust, FN provider ensures that the service will get paid and MU ensures that the FN provider is a legitimate and trusted provider.
- Agree on Session Profile: The two parties negotiate and agree on per-session features, such as what type of QoS is provided for which services. Agree on pricing and other billing related features.
- Agree on Session Security: The two parties negotiate and agree on mechanisms for protecting their traffic.

### 2.2.3 Policy Manager

Engaging parties use policies to govern the trust and negotiation. Policy Manager maintains the rules that meet the objective of every party in the negotiation and trust decision factors with the involving risk. At an abstract level, before conducting the negotiation each party prepares and specifies their own sets of policies that meet their own interests. The policies should include at least the following:

- Trust policy: the identification and credentials that

can be trusted.
- Identification, authentication, and authorization policies.
- Other policies governing per-session features, such as QoS, security, and billing settings.

## 2.3 Passport and Visa Protocol

This technique can be used when there is no roaming agreement between FN and MU's HN. It consists of two tokens Passport and Visa. The "Passport" is an authentication token that is issued by identity provider (IdP: HN or certificate authority) to MU in order to identify and verify MU identity. The Passport in itself does not grant any access, but provides a unique binding between an identifier and the subject. The "Visa" is an authorisation token that is granted to a MU via a FN. The Visa token can be used as an access control to ban individual users. The following are a set of protocols were developed to achieve the approach objective.

### 2.3.1 Passport Acquisition

This protocol will describe the MU registration process with HN (Passport issuer); by completing this protocol MU will receive a Passport (identification token). For a MU to get network services from a FN, this MU required to register with the HN to obtain a Passport.

The registration with the HN takes place offline. The registration with the HN occurs once and when completed, the HN issues a smart card (SC) to the MU. Every SC has a unique ID, which is combined with MU's biometric (such as finger print) to generate a symmetric master Key ($K_{MU-HN}$ denotes the shared key encryption between MU and HN). Key master is manually distributed, shared and stored on both parties' sides (HN and MU's SC) directly after issuing the SC. The HN's generate the MU's $Passport_{MU}^{HN}$ which is encrypted with the home network's public key (PK$_{HN}$ (X) denotes HN's public key encryption) and stored in the SC. The Passport is given as:

$$Passport_{MU}^{HN} = < PK_{HN}(id_{MU}, Pass_{No}, expiry, data, K_{MU-HN},$$
$$Sig_{HN}(id_{MU}, Pass_{No}, expiry, data, K_{MU-HN})) > \qquad (1)$$

In the Passport, $Sig_{HN}$ represents the digital signature of the Passport using the HN's private key which can be verified using the public key. Inside the Passport, "$id_{MU}$" is the mobile user's identity and "$Pass_{No}$" is the Passport number. The "expiry" field corresponds to the Passport expiry date. Finally, the field "data" consists of all other relevant information such as type of Passport, type of MU, MU name, MU date of birth, date of issue, place of issue, issuer ID, and issuer name etc.

### 2.3.2 Visa Acquisition

A MU will receive the required Visa (authorisation token) from the FN after completing the identification and verification process successfully. When the MU has his Passport (authentication token) in hand, the authentication process can be started with the FN in order to obtain

the required Visa. The protocol is demonstrated as follows:

$$MU \rightarrow FN : Passport_{MU}^{HN}, \{id_{FN}, r_{MU}, T_{MU}\}_{SK_{MU-HN}}, Pass_{No}, T_{MU},$$
$$Cert_{HN}, VisaReq_{FN}, r"_{MU"} \qquad (2)$$

$$FN \rightarrow HN: Passport_{MU}^{HN}, \{id_{FN}, r_{MU}, T_{MU}\}_{SK_{MU-HN}}$$
$$, T_{MU}, \ Cert_{FN}, T_{FN}, PK_{HN}(r_{FN}) \qquad (3)$$

$$HN \rightarrow FN: PK_{FN}(Pass_{No}, valid_{MU}, r_{MU}, r_{FN}, Sig_{HN}(Pass_{No}$$
$$, valid_{MU}, r_{MU}, r_{FN})), \{id_{FN}, valid_{FN}, r_{FN}, r_{MU}, T_{HN}\}_{SK_{MU-HN}} \qquad (4)$$

$$FN \rightarrow MU: Visa_{MU}^{FN}, \{id_{FN}, valid_{FN}, r_{FN}, r_{MU}, T_{HN}\}_{SK_{MU-HN}},$$
$$\{k_{MU-FN}\}_{SK_{MU-FN}}, r"_{FN"} \qquad (5)$$

$$SK_{MU-HN} = h(K_{MU-HN}, id_{MU}, id_{FN}) \qquad (6)$$

$$SK_{MU-FN} = h(Pass_{No}, id_{FN}, r_{MU}, r_{FN}, r"_{MU"}, r"_{FN"}) \qquad (7)$$

The protocol starts once a MU sends his *Passport* and the service Visa request in (2). The foreign network ID $id_{FN}$, MU's random number $r_{MU}$, and MU's timestamp are encrypted by the MU and HN session key $SK_{MU-HN}$. The generation of MU and HN session key $SK_{MU-HN}$ is illustrated in (6), where three factors named: shared master key $K_{MU-HN}$, MU' id, and HN's id are hashed using one way hash function $h(x)$. By computing this session key $SK_{MU-HN}$ a mutual authentication will be established between the MU and the HN. The FN's id is encrypted to enable the HN to verify the $id_{FN}$ with the one in the FN certificate and to make sure that it has not been modified by an attacker. The $r_{MU}$ used to authenticate the HN, and it is one of the factor of generating the MU and FN session key $SK_{MU-FN}$ in (7). By computing this session key $SK_{MU-FN}$ a mutual authentication will be established between the MU and the FN. The $Pass_{No}$ is used by the FN in the verification process with the $Pass_{No}$ that sent by the HN in (4). All the timestamps $T_{MU}, T_{FN}$, and $T_{HN}$ are used to stop reply attack. The HN's certificate $Cert_{HN}$ is sent to the FN for verification and establishing trust using certificate authority (CA). Another MU's random number $r"_{MU"}$ is sent to FN to be used as a factor in generating the MU and FN session key $SK_{MU-FN}$ in (7).

The FN forwards the encrypted Passport, $id_{FN}, r_{MU}, T_{MU}$ and adds its certificate $Cert_{FN}$, timestamp $T_{FN}$, random number encrypted by HN's public key ($PK_{HN}(r_{FN})$) to the HN in (3). The FN's certificate $Cert_{FN}$ is sent to the HN for verification and establishing trust using the certificate authority (CA). The FN's random number $r_{FN}$ used to authenticate the HN and it is one of the factor of generating the MU and FN session key $Sk_{MU-FN}$ in (7).

After receiving the message from the FN, the HN ensures if the timestamps of MU $T_{MU}$ and FN $T_{FN}$ are valid. If one of them is not valid, the HN replies with un-fresh session and terminate the request. Otherwise, the HN checks the validity of FN certificate $Cert_{FN}$ with the CA as a valid certificate as well as network provider. Then, the HN decrypts the Passport with its private key and then verifies the signature using the HN's public key. After the HN checks successfully the MU's Passport is genuine and valid, it get the shared key ($K_{MU-HN}$) and its relevant information such as the date of expiry. Then, the HN generates the session key ($SK_{MU-HN}$) for authentication and encryption between the MU and the HN. This session key will be used to decrypt the second part of the message $\{id_{FN}, r_{MU}, T_{MU}\}$. The HN compares the FN's ID in this massage with the one in the certificate to ensure the FN has not been changed. After verifying the FN, the HN decrypts the FN's random number by its private key. The HN gets the FN's random number $r_{FN}$, FN's ID from the certificate $id_{FN}$, the indicator of the validity of FN $valid_{FN}$, the MU's random number $r_{MU}$ and its timestamp $\{id_{FN}, valid_{FN}, r_{FN}, r_{MU}, T_{HN}\}$ and encrypt them with the MU-FN session key $Sk_{MU-HN}$. Also, as the HN authenticate the MU, the HN gets the MU's Passport ber $Pass_{No}$, the indicator of the validity of MU $valid_{MU}$, the MU's random number $r_{MU}$, and the FN's random number $r_{FN}$ ($Pass_{No}, valid_{MU}, r_{MU}, r_{FN}$). The HN then compute their digital signatures using its private key, then encrypt them using the FN's public key. Then, the HN put the FN authentication part and the MU authentication part in one massage and sends it to the FN.

Once the FN received the message from the HN, it decrypts its part using its private key and verifies it using the HN's public key. If the FN received the validity of the Passport and checks its random number, the Visa will be generated $Visa_{MU}^{FN}$ with the shared master key $K_{MU-FN}$. The FN encrypts the master key $K_{MU-FN}$ using the session key in (7) and sends it back to the MU with the Visa. Furthermore, the FN sends a new random number $r''_{FN}$ for generating the session key. The Visa is defined as:

$$Visa_{MU}^{FN} = < PK_{FN}(Pass_{No}, Visa_{No}, expiry, data, K_{MU-FN},$$
$$Sig_{FN}(Pass_{No}, Visa_{No}, expiry, data, K_{MU-FN})) > \qquad (8)$$

After the MU receives the authorisation message from the HN through FN, the MU generate the session key in (6) and decrypt the message. The HN's timestamps $T_{HN}$, random number $r_{MU}$, foreign network ID $id_{FN}$ correctness will be checked. If they were incorrect, the Visa will be rejected, and if they were verified the Visa will be kept for future service request. The MU compute the MU-FN session key $Sk_{MU-FN}$ to deycrpt the shared master key $K_{MU-FN}$. The Visa is encrypted with the FN's public key ($PK_{FN}(X)$ denotes the FN's public key encryption), which means that only the FN can decrypt it. The t $Pass_{No}$ is the Passport number of the MU. The Visa number $Visa_{No}$ is the unique identity of the Visa and the expiry is the Visa expiry date. The data field includes all detailed Visa information such as Visa type, number of access, duration of access, issuer place, issuer ID, issuer name, issued time, service type, service name, times of access. The signature of the FN $Sig_{FN}$ in the Visa is used to stop a forged Visa. Before the Visa is sent, the FN stores the Visa information for future verifications. The following is an example:

$$\{Pass_{No}; Visa_{No}; expiry; valid\} \qquad (9)$$

The $expiry$ is the Visa expiry date. The field valid is set to FALSE once a Visa is revoked, otherwise it is set to TRUE.

### 2.3.3 Mobile Service Provision

This protocol illustrates how a MU can be granted network services from a FN in secure manner. When the MU obtains a valid Visa, the MU can provide it to the FN for network service provision. The protocol is demonstrated as follows:

$$MU \rightarrow FN : SerReq, Visa_{MU}^{FN}, \{r_{MU''}, Visa_{No}\}_{SK'_{MU-FN}} \qquad (10)$$

$$FN \rightarrow MU : \{r_{FN''}, Pass_{No}\}_{SK'_{MU-FN}}, \{Service\}_{SK''_{MU-FN}} \qquad (11)$$

$$SK'_{MU-FN} = h(SK_{MU-FN}, Visa_{No}, Pass_{No}) \qquad (12)$$

$$SK''_{MU-FN} = h(SK'_{MU-FN}, K_{MU-FN}, r_{MU''}) \qquad (13)$$

$$SK'''_{MU-FN} = h(SK''_{MU-FN}, SK'_{MU-FN}, r_{FN''}) \qquad (14)$$

The protocol starts once the MU requests to validate the Visa to get access to the FN service. MU sends the Visa and encrypted its random number and Visa number with the first session key $SK'_{MU-FN}$. The first session key (demonstrated in (12)) is generated by using the hash function of three factors: last session key $K_{MU-FN}$, Passport and Visa number(received with the Visa). After the FN receives the service request, it decrypts the Visa with its private key and checks its validity with the FN's public key. Then, it compares this signature to the one received. If they are identical, the Visa is considered as valid. Then, it gets the number of the Visa $Visa_{No}$ and searches its database to see if the Visa is used for the first time. The $Visa_{No}$ is to prove to the FN the knowledge of the Visa contents. However, the FN have to computes the $SK'_{MU-FN}$ to verify $Visa_{No}$ and to get the new random number $r_{MU''}$. The new random number will be used to generate the second session key $SK''_{MU-FN}$ illustrated in (13). The second session key will be used by the FN to encrypt its random number $r_{FN''}$ and the Passport number $Pass_{No}$. Finally, the third session key will be generated $SK'''_{MU-FN}$ using the new FN random number $r_{FN''}$, the first and second session keys. By having the third session key in hand both parties know that mutual authentication has been realized and the service can be started.

### 2.3.4 Passport and Visa Revocation

This protocol will be used to stop stolen Passport or Visa. If a Passport or Visa is considered to be revoked (e.g., the mobile user's shared keys $K_{MU-HN}$ or $K_{MU-FN}$ expires or the MU notices the FN to revoke a Visa or the HN to revoke a Passport).

The Passport revocation can be illustrated as:

$$HN \rightarrow FN : PK_{FN} \begin{pmatrix} Pass_{No}, RevOke, \\ Sig_{HN}(Pass_{No}, RevOke) \end{pmatrix} \qquad (15)$$

The protocol starts when the HN sends the RevOke message (which has been encrypted with the FN's public key and signed by the HN's private key) to the corresponding FN. The FN decrypts the message with its private key and verifies the signature with the HN's public key. The FN checks if the Passport number $Pass_{No}$ is already stored. If not, it means that there is no Visa issued with this Passport number. If it was stored, it stores the revoked Passport information and updates the status of the Visa as RevOke.

The Visa revocation can be illustrated as:

$$MU \rightarrow FN: \left\{ \begin{array}{c} Pass_{No}, Visa_{No}, RevOke, \\ \{Pass_{No}, Visa_{No}, RevOke\}_{SK'_{MU-FN}} \end{array} \right\}_{SK_{MU-FN}} \quad (16)$$

When FN receives a RevOke message from MU, the FN decrypts the message with the last session key $SK_{MU-FN}$ then verifies that by decrypting the other part of the message with the first session key $SK'_{MU-FN}$ (illustrated in (12)).The FN updates the status of the Visa as RevOke. Once a MU requests network services, the FN checks if the Visa was revoked. If it is revoked the service request will be rejected.

## 3 SYSTEM SECURITY ANALYSIS

### 3.1 Logical proof by BAN Authentication logic

The objective of our Passport-Visa mechanism is to provide authentication services for all parties engaged in the communication. Authentication protocols rely on network communication [8] and any failed communication will allow an opponent to pretend to be an lawful node in running the protocol. Authentication logics are tools that can be employed to assess the trust relationships in authentication protocols. The BAN logic will be utilized [9] to examine and verify the security of the Passport and Visa protocols. The assumptions of initial beliefs of the engaging parties are illustrated below in order to examine our protocol logic:

MU believes fresh ($\mathbf{r_{MU}}$) , FN believes fresh ($\mathbf{r_{MU}}$)

MU believes fresh ($\mathbf{r_{MU'}}$) , FN believes fresh ($\mathbf{r_{MU'}}$)

MU believes fresh ($\mathbf{r''_{MU''}}$) , FN believes fresh ($\mathbf{r''_{MU''}}$)

MU believes fresh ($\mathbf{r_{FN}}$) , FN believes fresh ($\mathbf{r_{FN}}$)

MU believes fresh ($\mathbf{r_{FN'}}$) , FN believes fresh ($\mathbf{r_{FN'}}$)

MU believes fresh ($\mathbf{r''_{FN''}}$) , FN believes fresh ($\mathbf{r''_{FN''}}$)

HN believes fresh ($\mathbf{r_{MU}}$) , HN believes fresh ($\mathbf{r_{FN}}$)

MU believes $\mathbf{MU} \xrightarrow{\mathbf{K}} \mathbf{HN}$ , HN believes $\mathbf{MU} \xrightarrow{\mathbf{K}} \mathbf{HN}$

MU believes $\mathbf{MU} \xrightarrow{\mathbf{SK}} \mathbf{HN}$ , HN believes $\mathbf{MU} \xrightarrow{\mathbf{SK}} \mathbf{HN}$

MU believes $\mathbf{MU} \xrightarrow{\mathbf{K}} \mathbf{FN}$ , FN believes $\mathbf{MU} \xrightarrow{\mathbf{K}} \mathbf{FN}$

MU believes $\mathbf{MU} \xrightarrow{\mathbf{SK}} \mathbf{FN}$ , FN believes $\mathbf{MU} \xrightarrow{\mathbf{SK}} \mathbf{FN}$

MU believes $\mathbf{MU} \xrightarrow{\mathbf{SK''}} \mathbf{FN}$ , FN believes $\mathbf{MU} \xrightarrow{\mathbf{SK''}} \mathbf{FN}$

MU believes $\mathbf{MU} \xrightarrow{\mathbf{SK'''}} \mathbf{FN}$ , FN believes $\mathbf{MU} \xrightarrow{\mathbf{SK'''}} \mathbf{FN}$

MU believes (FN controls $\mathbf{MU} \xrightarrow{\mathbf{K}} \mathbf{FN}$)

FN believes fresh $\mathbf{MU} \xrightarrow{\mathbf{K}} \mathbf{FN}$

In the demonstrated believes, the three engaging parties believe the freshness of $r_{MU}$. Both the FN and the MU believe the freshness of $r_{FN}$ and $r_{MU'}$. Together the MU and the FN believe the freshness of the secret key $MU \xrightarrow{K} FN$ and trust it is generated by the HN. Also, each party trusts its own shared key.

The idealized form of the Visa acquisition and service provision protocols is below:

M1: $MU \rightarrow FN: PK_{HN}\left(\left\{MU \xrightarrow{K} HN\right\}\right), \{r_{MU}\}_{SK_{MU-HN}}$ (17)

M2: $FN \rightarrow HN: PK_{HN}\left(\left\{MU \xrightarrow{K} HN\right\}\right)$

$, \{r_{MU}\}_{SK_{MU-HN}}, PK_{HN}(r_{FN})$ (18)

M3: $\rightarrow FN: PK_{FN}(r_{MU}, r_{FN}, Sig_{HN}(r_{MU}, r_{FN}))$,

$\{r_{FN}, r_{MU}\}_{SK_{MU-HN}}$ (19)

M4: $FN \rightarrow MU: PK_{FN}\left(\left\{MU \xrightarrow{K} FN\right\}\right), \{r_{MU}, r_{FN}\}_{SK_{MU-HN}}$,

$\left\{MU \xrightarrow{K} FN\right\}_{SK_{MU-FN}}$ (20)

M5: $MU \rightarrow FN: PK_{FN}\left(\left\{MU \xrightarrow{K} FN\right\}\right), \{r_{MU''}\}_{SK'_{MU-FN}}$ (21)

M6: $FN \rightarrow MU: \{r_{FN''}\}_{SK''_{MU-FN}}, MU \xrightarrow{SK'''} FN$

In the M1 message, the foreign network FN cannot sees $PK_{HN}\left(\left\{MU \xrightarrow{K} HN\right\}\right)$ and $(r_{MU})_{SK_{MU-HN}}$, instead it forwards them to the HN. When receiving the M2, the HN sees $PK_{HN}\left(\left\{MU \xrightarrow{K} HN\right\}\right)$ $, \{r_{MU}\}_{SK_{MU-HN}}$ and $PK_{HN}(r_{FN})$. So the HN can read the MU's random number, the shared key , and get the random number $r_{FN}$ of the FN after decrypt the message. As the HN believes $MU \xrightarrow{SK} HN$, the HN believes the MU said $MU \xrightarrow{SK} HN$ and realize the HN believes MU said ( $r_{MU}$). So the HN believes that it is the MU, not anyone else, who sent the message. The initial belief the HN believes fresh ($r_{MU}$), so the HN believes that the MU believes fresh ($r_{MU}$).

Similarly, when receiving M3, we have the FN believes $r_{MU}$, and $r_{FN}$. So both the HN and the FN know the secrets $r_{MU}$, and $r_{FN}$.

While MU receives M4, with the result that MU believes $MU \xrightarrow{K} FN$. Consequently, we can have the following result: MU believes FN said $r_{MU}$ and $r_{FN}$. Since the FN believes $MU \xrightarrow{SK} FN$, we obtain the MU believes $MU \xrightarrow{SK} FN$. This means that MU believes that $MU \xrightarrow{K} FN$ is the shared secret key between the MU and the FN.

At this stage, MU and FN are mutually authenticated. The last message, M5, is specifically used for the new session key agreement. In this message the FN believes fresh $r_{MU''}$ and the FN sees $SK'_{MU-FN}$ the first session key. Finally, it is M6 where both the FN and the MU believes $MU \xrightarrow{SK''} FN$ and $MU \xrightarrow{SK'''} FN$ which are the second and third session keys respectively.

In the following we will summarize the previous formal analysis with natural language as follows. Since the Passport is encrypted by the HN and signed by the HN, only the HN can decrypted through its private key and verify it using its public key. After that, the HN can get the MU's shared key and compute the session key by hashing with the ID of both the MU and the FN. This session key will decrypt the random number $r_{MU}$. As the Visa is encrypted by the FN's public key, only FN can decrypt it. Once MU receives the shared key $K_{MU-FN}$ from the FN, both the MU and the FN believe that $K_{MU-FN}$ is a shared secret key between them. They trust each other and use this key to generate their session keys. When the FN receives this Visa, it can get the shared key $K_{MU-FN}$ stored in the Visa and use the random numbers, Passport and Visa numbers to computes the first, second and third session keys. Both the MU and the FN contribute a random number and generate alone the new session key (SK) for the use in the current session.

## 3.2  Security Analysis

As a limitation of the BAN logic that it cannot deal with all security flaws [10], we are to consider some of the potential attacks in order to analyse security of the proposed protocol.

1) Forge Passport-Visa: a valid signature cannot be generated by external attackers in the name of the HN or FN. So it is impossible for them to fabricate or fake Passport or Visa. As the MU's Visa can only be verified by the FN, therefore only the FN can know the secret shared key $K_{MU-FN}$ of the MU and retrieve the random numbers. Without the shared key and the random numbers the attacker cannot generate the session key $SK_{MU-FN}$.

2) Mutual authentication: In the mobile service provision phase, the MU sends a message that consist of two part: Visa, and the encrypted new random number $r_{MU''}$. The FN decrepts the Visa with it is public key and get the shared key. Also as the FN signed the Visa, so it can check the validation of the Visa. The FN use the previous session key with $Pass_{No}, Visa_{No}$ to generate the first session key which will be used to decrypt the second part of the message and get the new random number. The shared master key with the first session key and $r_{MU''}$ will be used to generate the second session key. By decrypting the FN message, the MU can get the FN's random number. Now, both parties are able to generate the third session key and mutual authenticate each other.

3) Replay and man-in-the-middle attacks: An attacker may sniff a valid Visa, however the $K_{MU-FN}$, $Pass_{No}$, and $Visa_{No}$ cannot be got as they are encrypted in the Visa. The only party that can get the $K_{MU-FN}$, $Pass_{No}$, and $Visa_{No}$ from the Visa is the FN.

4) Key freshness: only the MU and the FN know the shared key $K_{MU-FN}$. In addition, it not used in any communication. In every service provision the new session key is generated, but it is valid just in that current session. This key is established by contributing the random numbers of both the MU and the FN. So the key freshness is guaranteed.

## 4  RELATED WORK

There are number related works in the area of ubiquitous mobile access authentication. Their strengths and limitations will be discussed.

There are a number of proposals that based on the ticket model [8, 11-17]. One of them is Kerberos [18-22] the famous ticket model which used for network authentication and control access to network services. In Kerberos, tickets have a limited life time which means it should be used for the current session only. This will involve more loads in the ticket server. Also, the key distribution centre (KDC) in Kerberos maintains all the shared keys for every registered party. In our proposal, the Visa can be used multiple times to access FN unless the Visa is expired or met the specific access number. The security of the Visa can be maintained by generating new fresh session key for every session using the random numbers. Also, both HN and FN do not require to store the shared key of MUs as the can be obtained from the Passport and the Visa.

In this inspiring paper [8] Lei, Quintero and Pierre presented a reusable tickets for accessing mobile services. In their proposal a lightweight computational symmetric keys are used on the mobile device side to support the limited capabilities of MD. The major difference between this work and our proposal is that, in their protocol FN does not have a control over granting the authorisation token, as the tickets are approved by the ticket server. Therefore, their approach will not work in case there is no service level agreement with the potential FN. For example, a MU wants to access network services from new FN that not yet establish service agreement with ticket server or the FN is not large enough to be approved by the ticket server. In the Passport-Visa approach the FN has a full control wither to grant an authorisation access (Visa) to this individual MU or not. This approach give more freedom for MU in choosing the service provider based on direct negotiation of the services, identification (Passport), and authorisation (Visa). In this paper the FN is an independent party and it does not require service level agreement with MU's HN to provide the network services.

Butty´an and Hubaux [23] have proposed a model based on the introduction of customer care agencies (CCA) and a ticket based mechanism for all kinds of mobile services. The goal of their model is to enable MUs to choose their service providers in a more flexible way, handle payments on behalf of the user, and take care of protecting the user's privacy by the assistance of customer care agency. The problem with this model is similar to the previous work where the CCA act as a broker where every service provider should establish a service agreement to be accessible to MUs. This solution does not support the open market environment as MUs depend on CCA to access network providers, and there is no direct negotiation between MUs and FN providers. Another issue that the CCA require the MU to generate session key using Diffie-Hellman scheme. It is difficult to be computed by a limited resources device.

Akyildiz and Mohanty [24] have proposed an Architecture for ubiquitous Mobile Communications (AMC). Their aim is to provide ubiquitous high-data rate services to MUs by integrating heterogeneous wireless systems. AMC eliminates the need for direct roaming agreements among network providers by using a third party network interoperating agent (NIA). The NIA acts as a broker, and it requires network providers to have pre-established roaming agreements. This solution does not support the open market environment as MUs depend on NIA to access network providers, and there is no direct negotiation between MUs and FN providers.

Droma and Ganchev [25-27] have proposed a Consumer-centric Business Model (CBM) for wireless services. They argue that their model is a better alternative to the subscriber based model (SBM). In the CBM model, entities should have business agreement only with the third-party authentication, authorization, and accounting (3P-AAA) service provisions. The 3P-AAA-SPs are independent entities and not wireless access network providers (ANPs). This research has some common aims with this project research such as ubiquitous wireless access and more open marketplace. However, the 3P-AAA-SP works as a broker and it requires network providers to have pre-established service agreement.

Shi et al. [28] have introduced Service Agent (SA) to the WLAN/cellular integrated network architecture to improve service flexibility and deal with the roaming agreement issue when the number of WLAN operators is large. The SA provides cellular network and WLAN with one-for-all roaming agreement so that one-to-one roaming agreements are no longer needed. In their proposed service model, the MU does not have to be a customer of any physical network operator. The SA can provide cellular/WLAN integrated service itself. This approach has the same limitation of the broker model as well as it is dependent on Wi-Fi and cellular network technologies only.

Matsunaga et al. [29] have proposed a single sign-on (SSO) authentication architecture that confederates WLAN service providers through trusted IdPs. They argue that the dynamic selection of authentication method and IdP will play a key role in confederating public wireless LAN service providers under different trust levels and with alternative authentication schemes. In their implementation they used two different industry standard single sign-on authentication schemes in public wireless LANs: RADIUS and Liberty Architecture. A client-side policy engine enables the user to select which of the alternate single sign on authentication schemes to use. The first limitation in this approach is the dependence in roaming agreement between network providers and IdPs, which may limit the MU roaming freedom. The second limitation is the dependency on a single wireless technology. Lastly, it is limited to web-based authentication using cookies [30].

Sastry et al. [31, 32] proposed to keep the Wi-Fi HN as the actual network provider even in the visiting network by creating a tunnel between FNs and HN to answer all the MU service requests directly by the HN. Their aim is to provide a roaming across wireless LAN that eliminates the security concerns of a FN. However, the MU is restricted to roam across limited FNs that have roaming agreements with its HN, which limits the freedom of selecting the most appropriate network. This approach also is not efficient as it forward the MU traffic to the far away HN which involve high latency and increase the network traffic.

Bahl et al. [33] have proposed the CHOICE network architecture and its underlying protocol PANS (Protocol for Authorization and Negotiation of Services). Their goal is to globally authenticate users and securely connects them to the Internet via a high-speed wireless LAN. In their work they use Microsoft Passport as their global authenticator. To gain access to the network, the user should authenticate himself/herself with the global authenticator obtaining a key from the PANS authorizer. However, this approach is wireless technology dependent. Also, the global authenticator act as a broker which requires all WLAN providers to have a pre-established service agreement. Furthermore, using a simple username and a password as in Microsoft Passport is a weak authentication.

Chakravorty et al. [34] proposed a mobile bazaar (MoB) , an open market architecture for collaborative wide-area wireless services by using reputation management and third party accounting and billing. Their approach is based on short-term transient access network resource reselling by the network's subscribers to other users using an ad hoc network type solution. Their aim is to provide the MU with network access and freedom to choose a better connection (high bandwidth) in a FN domain by trading with FN users. An available idle terminal may act as an access node (i.e., effectively as an ad hoc wireless router) to provide access, directly or via a multihop link, to wireless communications resources such as a 3G cellular network, and receive payment for this service. The limitation of this approach is the dependency on FN's users availability in trading and accessing the network.

Fu et al. [35] have proposed in fly partnership negotiations to achieve spontaneous and dynamic roaming agreements using policy based negotiations for heterogeneous network providers. Their approach aims to elimi-nate manually set up pre-established formal roaming agreements, which is a costly and time-consuming process. They argue for the need to establish on the fly roaming agreement to optimize the network providers' cooperation. However, this approach is based on the centralized model, which depends on the HN for accessing FNs. And there is no direct negotiation with the MU.

Shin et al. [30, 36] argue that centralized authentication approaches are inefficient as the HN participates in each authentication process, causing high latency. They have proposed a chained method of distributed authentication for internetwork. The role of HN authentication has been limited to the first visited network, where the rest relies on the previous visited network for authentication. This approach relies on the collaboration between adjacent networks and the level of trust and requires service agreement between them. Also, there is no direct negotiation between a MU and FNs.

Tuladhar et al.[3] have proposed proof tokens authentication architecture and protocol. It is similar to the previous approach (Chained Authentication), as it reduces the need for HN authentication by making use of the previous trusted visited network to authenticate the MU. In their approach, they tried to solve two problems. The first problem is the limited roaming agreement of HN with FNs, and they propose to allow MUs to access the partners of previously visited networks by that MU. The second problem is authentication delay, which they identified as a major cause for high latency, and propose the collaboration between adjacent networks. However, this approach still relies on roaming agreement for authentication, and does not support a direct negotiation with the MU.

## 5 CONCLUSION

This paper argued for the need of a flexible way to authenticate mobile users in ubiquitous wireless access environment. We proposed a roaming agreement-less approach as a practical solution to provide MU with a flexible way to authenticate themselves outside the HN partners' domain with direct negotiation with FN provider. The FNs have full control over the authorisation process. The aim of this paper is to enable MU to access more wireless FN providers without relying on HN's roaming agreements. The security analysis indicates that our Passport and Visa proposals provide a secure authentication mechanism for ubiquitous wireless access environment. Our proposed technique will provide MU with more network services from FN domains in a secure manner.

As further work, we are currently working to enhance the Passport and Visa protocols security and usability. A very promising technique we are going to analyse to be adapted in our approach is the limited-used key theory [37-39]. The main idea behind this theory is one-time used symmetric cryptographic keys that improve the security of the cryptographic system significantly. The dynamic keys system encrypt every message with a different key, therefore even if the attacker finds out the key for one packet it still cannot read the whole message as the attacker needs to find out the other packets keys.

## REFERENCES

[1] GSM Association, "20 Facts for 20 Years of Mobile Communications," in,http://www.gsmtwenty.com/20facts.pdf,Cellular News, Mobile Phone Subscribers Pass 4 Billion Mark, Cellular News, http://www.cellular-news.com/story/35298.php.

[2] E. Gustafsson and A. Jonsson, "Always best connected," IEEE Wireless Communications, vol. 10, pp. 49-55, 2003.

[3] S. Tuladhar, C. Caicedo, and J. Joshi, "Inter-Domain Authentication for Seamless Roaming in Heterogeneous Wireless Networks," 2008, pp. 249-255.

[4] Maria Thurrell and M. Awaken, "tripwolf and FON Launch Competition to Promote Universal Free WiFi for Global Travelers." vol. 2009: Free Press Release, 2008.

[5] A. Jsang, C. Keser, and T. Dimitrakos, "Can we manage trust," 2005.

[6] S. Ruohomaa, "Trust management for inter-enterprise collaborations," 2007.

[7] Z. Fu, M. Shin, J. C. Strassner, N. Jain, V. Ram, and W. A. Arbaugh, "AAA for Spontaneous Roaming Agreements in Heterogeneous Wireless Networks," Lecture Notes in Computer Science, vol. 4610, p. 489, 2007.

[8] Y. Lei, A. Quintero, and S. Pierre, "Mobile services access and payment through reusable tickets," Computer Communications, 2008.

[9] M. Burrows, M. Abadi, and R. Needham, "A logic of authentication," ACM Transactions on Computer Systems (TOCS), vol. 8, pp. 18-36, 1990.

[10] C. Boyd and W. Mao, "On a limitation of BAN logic," Lecture Notes in Computer Science, vol. 765, pp. 240-247, 1994.

[11] B. Patel and J. Crowcroft, "Ticket based service access for the mobile user," 1997, pp. 223-233.

[12] H. Wang, J. Cao, and Y. Zhang, "Ticket-based service access scheme for mobile users," 2002, pp. 285-292.

[13] B. Lee, T. Kim, and S. Kang, "Ticket based authentication and payment protocol for mobiletelecommunications systems," 2001, pp. 218-221.

[14] Y. Chen, C. Chen, and J. Jan, "A mobile ticket system based on personal trusted device," Wireless Personal Communications, vol. 40, pp. 569-578, 2007.

[15] H. Wang, X. Huang, and G. Dodda, "Ticket-based mobile commerce system and its implementation," 2006, pp. 119-122.

[16] H. Wang, Y. Zhang, J. Cao, and Y. Kambayahsi, "A global ticket-based access scheme for mobile users," Information Systems Frontiers, vol. 6, pp. 35-46, 2004.

[17] H. Wang, Y. Zhang, J. Cao, and V. Varadharajan, "Achieving secure and flexible m-services through tickets," IEEE Transactions on Systems, Man, and Cybernetics, Part A: Systems and Humans, vol. 33, pp. 697-708, 2003.

[18] S. P. Miller, B. C. Neuman, J. I. Schiller, and J. H. Saltzer, "Kerberos authentication and authorization system," In Project Athena Technical Plan, 1987.

[19] C. Neuman, T. Yu, S. Hartman, and K. Raeburn, "The Kerberos network authentication service (v5)," ISI, 1993.

[20] B. C. Neuman and T. Ts'o, "Kerberos: An authentication service for computer networks," IEEE Communications Magazine, vol. 32, pp. 33-38, 1994.

[21] J. Kohl, B. C. Neuman, and J. Steiner, "The Kerberos network authentication service," 1991.

[22] J. T. Kohl, B. C. Neuman, and T. Y. Ts'o, The evolution of the Kerberos authentication service: University of Southern California, Information Sciences Institute, 1994.

[23] L. Buttyan and J. Hubaux, "Accountable anonymous access to services in mobile communicationsystems," 1999, pp. 384-389.

[24] I. Akyildiz, S. Mohanty, and J. Xie, "A ubiquitous mobile communication architecture for next-generation heterogeneous wireless systems," IEEE Communications Magazine, vol. 43, pp. S29-S36, 2005.

[25] M. O'Droma and I. Ganchev, "Toward a ubiquitous consumer wireless world," IEEE Wireless Communications, vol. 14, pp. 52-63, 2007.

[26] I. Ganchev, M. O'Droma, and N. Wang, "Consumer-Oriented Incoming Call Connection Service for a Ubiquitous Consumer Wireless World," Wireless Personal Communications, vol. 50, pp. 115-131, 2009.

[27] M. O'Droma and I. Ganchev, "Strategic innovations through NGN standardisation for a Ubiquitous Consumer Wireless World," 2008, pp. 135-142.

[28] M. Shi, H. Rutagemwa, X. Shen, J. W. Mark, and A. Saleh, "A Service-Agent-Based Roaming Architecture for WLAN/Cellular Integrated Networks," IEEE Transactions on Vehicular Technology, vol. 56, pp. 3168-3181, 2007.

[29] Y. Matsunaga, A. Merino, T. Suzuki, and R. Katz, "Secure authentication system for public WLAN roaming," 2003, pp. 113-121.

[30] M. Shin, J. Ma, and W. Arbaugh, "The Design of Efficient Internetwork Authentication for Ubiquitous Wireless Communications," Network, vol. 3, p. 1, 2004.

[31] N. Sastry, J. Crowcroft, and K. Sollins, "Architecting citywide ubiquitous wi-fi access," 2007.

[32] M. Manulis, D. Leroy, F. Koeune, O. Bonaventure, and J. Quisquater, "Authenticated Wireless Roaming via Tunnels: Making Mobile Guests Feel at Home?," 2009.

[33] P. Bahl, A. Balachandran, and S. Venkatachary, "Secure wireless internet access in public places," 2001, pp. 3271–3275.

[34] R. Chakravorty, S. Agarwal, S. Banerjee, and I. Pratt, "MoB: a mobile bazaar for wide-area wireless services," 2005, pp. 228-242.

[35] Z. Fu, M. Shin, J. Strassner, N. Jain, V. Ram, and W. Arbaugh, "AAA for Spontaneous Roaming Agreements in Heterogeneous Wireless Networks," Lecture Notes in Computer Science, vol. 4610, p. 489, 2007.

[36] M. Shin, J. Ma, A. Mishra, and W. A. Arbaugh, "Wireless network security and interworking," Proceedings of the IEEE, vol. 94, pp. 455-466, 2006.

[37] A. Rubin and R. Wright, "Off-line generation of limited-use credit card numbers," Lecture Notes in Computer Science, vol. 2339, pp. 196-209, 2001.

[38] S. Kungpisdan, P. Le, and B. Srinivasan, "A limited-used key generation scheme for internet transactions," Lecture Notes in Computer Science, vol. 3325, pp. 302-316, 2005.

[39] X. Wu, P. Le, and B. Srinivasan, "Dynamic Keys Based Sensitive Information System," in The 9th International Conference for Young Computer Scientists (ICYCS 2008), Zhang Jia Jie, Hunan, China, 2008, pp. 1895-1901.

**Abdullah Almuhaideb** received the B.S (Hons) degree in Computer Information System from King Faisal University in 2003 and M.S degree in Network Computing from Monash University in 2007. He is currently working towards the PhD degree at Monash University in Caulfield School of Information Technology. Abdullah's research interests include: Ubiquitous Wireless Access, Mobile Security, Authentication & Identification. Also, he is a lecturer at the Computer Networks Department, King Faisal University.
Previous publications include:
- "Beyond Fixed Key Size: Classification toward a Balance between Security and Performance", AINA 2010: The International Conference on Advanced Information Networking and Applications, April 20-23 2010, Perth, Australia, In press.
- "Extended Abstract: an Adaptable Multi-Level Security based on different Algorithms Key Sizes for Mobile Devices", SIC 2009: the 3rd Saudi International Conference, June 5-6 2009, University of Surrey, Guildford, UK.
- "Comparative Efficiency and Implementation Issues of Itinerant Agent Language on Different Agent Platforms ", AT2AI-6: From Agent Theory to Agent Implementation Workshop in the scope of AAMAS 2008 ( The 7th International Conference on Autonomous Agents and Multiagent Systems), May 12-16 2008, Cascais, Portugal.

**Mohammed Alhabeeb** received the B.S (Hons) degree in Computer Information System from King Saud University in 1999 and M.S degree in Network Computing from Monash University in 2007. He is currently working towards the PhD degree at Monash University in Caulfield School of Information Technology. Mohammed's research interests include: denial of services, information security, and security analysis. Also, he is a project manager at the National Information Centre, Ministry of Interior in Saudi Arabia.
Previous publications include:
- "Holistic Approach for Critical System Security: Flooding Prevention and Malicious Packet Stopping", Journal of Telecommunications, Volume 1, Issue 1, pp14-24, February 2010.
- "Information Security Threat Classification Pyramid", FINA: The Sixth International Symposium on Frontiers of Information Systems and Network Applications (FINA), April 20-23 2010, Perth, Australia, In press.

**Dr Phu Dung Le** is currently working at School of Information Technology. Dr Le's main research interests are: Image and Video Quality Measure and Compression, Intelligent Mobile Agents, Security in Quantum Computing Age. He used to teach Data Communication, Operating System, Computer Architecture, Information Retrieval and Unix Programming. He has also researched in Mobile Computing, Distributed Migration. Currently he is lecturing network security and advanced network security in addition to supervising PhD students.
Previous publications include:
- "A Tool for Migration to Support Resource and Load Sharing in Heterogeneous Environments", Proceedings of the International Conference on Networks, pp. 83-87, Feb 1996
- "A Limited-used Key Generation Scheme for Internet Transations". Information Security Applications, Vol. 3325, pp 302-316, Lecture Notes in Computer Science, ISBN: 3-540-24015-2, Korea, 2005
- "The Design and Implementation of a Smart Phone Payment System", IEEE Proceedings of Information Technology: New Generations, pp. 458-463, USA 2006

**Bala Srinivasan** is a Professor of Information Technology in the Faculty of Information Technology, Monash University, Melbourne, Australia. He has authored and jointly edited 6 technical books and authored and co-authored more than 200 international refereed publications in journals and conferences in the areas of Databases, Multimedia Retrieval Systems, Distributed and Mobile Computing and Data Mining. He has successfully supervised 28 research students of which 15 of them are PhDs and, his contribution to research supervision has been recognised by Monash University by awarding him the Vice-Chancellors medal for excellence. He is a founding chairman of the Australiasian database conference which is now being held annually. He holds a Bachelor of Engineering Honours degree in Electronics and Communication Engineering from University of Madras, a Masters and a Ph.D, both in Computer Science from Indian Institute of Technology, Kanpur. Currently he is in the editorial board of two international journals and program committee member of nearly dozen international conferences.